\documentclass[12pt]{article}
\usepackage{amsmath,epsf,amssymb,latexsym,cite,xy,curves}
\xyoption{matrix}\xyoption{arrow}

\newif\iffigs\figstrue

\DeclareFontFamily{U}{rsf}{}
\DeclareFontShape{U}{rsf}{m}{n}{
  <5> <6> rsfs5 <7> <8> <9> rsfs7 <10-> rsfs10}{}
\DeclareMathAlphabet\Scr{U}{rsf}{m}{n}

\DeclareMathAlphabet\mathbi{U}{cmr}{bx}{it}

\def\pplogo{\vbox{\kern-\headheight\kern -29pt
\halign{##&##\hfil\cr&{
\ppnumber}\cr\rule{0pt}{2.5ex}&\ppdate\cr}
}}
\makeatletter
\def\ps@firstpage{\ps@empty \def\@oddhead{\hss\pplogo}%
  \let\@evenhead\@oddhead 
}
\def\maketitle{\par
 \begingroup
 \def\thefootnote{\fnsymbol{footnote}}
 \def\@makefnmark{\hbox{$^{\@thefnmark}$\hss}}
 \if@twocolumn
 \twocolumn[\@maketitle]
 \else \newpage
 \global\@topnum\z@ \@maketitle \fi\thispagestyle{firstpage}\@thanks
 \endgroup
 \setcounter{footnote}{0}
 \let\maketitle\relax
 \let\@maketitle\relax
 \gdef\@thanks{}\gdef\@author{}\gdef\@title{}\let\thanks\relax}
\makeatother

\def\O{\Scr{O}}
\def\C{{\mathbb C}}
\def\P{{\mathbb P}}

\def\R{{\mathbb R}}
\def\Z{{\mathbb Z}}

\def\Hom{\operatorname{Hom}}

\def\Ext{\operatorname{Ext}}

\def\SU{\operatorname{SU}}

\def\Cone{\operatorname{Cone}}

\def\ch{\operatorname{\mathit{ch}}}
\def\td{\operatorname{\mathit{td}}}

\def\p{\partial}

\def\CY{Calabi--Yau}

\def\cE{{\Scr E}}
\def\cF{{\Scr F}}
\def\cX{{\Scr X}}
\def\cY{{\Scr Y}}

\def\DC{\mathbf{D}}

\def\ff#1#2{{\textstyle\frac{#1}{#2}}}

\def\labto#1{\mathrel{\mathop\to^{#1}}}

\begin{document}
\setcounter{page}0
\def\ppnumber{\vbox{\baselineskip14pt
\hbox{DUKE-CGTP-02-10}
\hbox{hep-th/0211121}}}
\def\ppdate{November 2002} \date{}

\title{\LARGE Solitons in Seiberg--Witten Theory\\ and D-branes in 
the Derived Category\\[10mm]} 
\author{
Paul S.~Aspinwall and Robert L.~Karp\\[4mm]
\normalsize Center for Geometry and Theoretical Physics \\
\normalsize Box 90318 \\
\normalsize Duke University \\
\normalsize Durham, NC 27708-0318 \\[8mm]
}

{\hfuzz=10cm\maketitle}

\def\Large{\large}
\def\LARGE{\large\bf}

\vskip 1cm

\begin{abstract}
We analyze the ``geometric engineering'' limit of a type II string on
a suitable \CY\ threefold to obtain an $N=2$ pure $\SU(2)$ gauge
theory. The derived category picture together with $\Pi$-stability of
B-branes beautifully reproduces the known spectrum of BPS solitons in
this case in a very explicit way. Much of the analysis is particularly
easy since it can be reduced to questions about the derived
category of $\P^1$.
\end{abstract}

\vfil\break


\section{Introduction}    \label{s:intro}

Since the work of Seiberg and Witten \cite{SW:I} several programs have
emerged allowing one to compute the stable spectrum of BPS solitons at any
given point in the moduli space. A knowledge of central charges allows
one to compute the location of walls of stability where solitons {\em
might} decay but one must work harder to show that a given soliton really
does decay. This was first done in \cite{BF:su2}, indirectly, by using
global properties of the moduli space.

One may compactify a type II string on a \CY\ threefold  $X$ to obtain a
theory in four dimensions with a nonperturbative nonabelian gauge
symmetry. The systematics of obtaining a particular gauge group with a
particular matter content are well understood \cite{AKM:lcy}.  By going to a
particular corner of the moduli space \cite{KKL:limit} one can then
decouple gravity to obtain a supersymmetric Yang--Mills field theory. One
says the field theory has been ``geometrically engineered''
\cite{KKV:geng} from a type II string theory.

One would then like to obtain the spectrum of BPS solitons of the Yang--Mills
theory from a knowledge of the BPS states in the type II string theory
--- namely BPS D-branes. BPS D-branes occur in two types on a \CY\
threefold. The A-branes may be na\"\i vely pictured as special
Lagrangian 3-manifolds in $X$, while B-branes may be na\"\i vely
pictured as holomorphic subspaces of $X$. The fact that neither of
these statements is quite right makes D-branes on \CY\ threefolds a
very interesting area of current research.

In \cite{KLM:hSW} (see also \cite{Lc:SWintro}) a method of analyzing
solitons as A-branes in this context was given.  In recent years many
other approaches to analyzing the BPS spectrum have appeared, including
\cite{RSVV:stab,AN:webs,Dnf:Dstab,DGR:Dquin} to name but a few.  In
this paper we will directly analyze the soliton spectrum yet another
way, by using B-branes, the derived category of coherent sheaves and
$\Pi$-stability. We justify this in several ways:
\begin{itemize}
\item We believe that ultimately this
provides a relatively simple way of determining soliton spectra, even
though one has to first learn the machinery of the derived category.
\item One can view this
paper as a test of the idea of $\Pi$-stability, which it passes
beautifully. 
\item This is probably the easiest example of $\Pi$-stability to
  understand completely.
\item One might argue that the degree of rigour in
the B-brane picture compares very favourably with most of the other
methods. For example, at this point in time it is probably safe to say
that B-branes are better-understood than A-branes, since the
topological A-model is subject to world-sheet instanton corrections.
\end{itemize}

Recent progress in the derived category picture
\cite{Kon:mir,Doug:DC,Laz:DC,AL:DC,Dia:DC} appears to promise a
substantially complete understanding of B-branes in the limit of weak
string coupling. In particular it incorporates all
$\alpha'$-corrections.  Since we wish to analyze the stability of
solitons, the key question for us to analyze will concern the {\em
stability\/} of objects in the derived category. This is the subject
of ``$\Pi$-stability''. Criteria for $\Pi$-stability have been
discussed in \cite{DFR:stab,DFR:orbifold,Doug:Dgeom,AD:Dstab} with
further examples discussed in \cite{me:point,AKH:m0}.  The technique
consists of starting at large radius limit, where one may use the
classical picture of B-branes as holomorphic submanifolds, and then
following the relative ``gradings'' of the stable branes as one moves
along paths in the moduli space.

This is quite a formidable computation in general, but it has been pursued
on some cases (see \cite{DDG:wrap,AD:Dstab,me:point} for example). In
the model studied in this paper, the \CY\ threefold is a K3-fibration
over a $\P^1$. Much of the analysis can be described using the derived
category of $\P^1$, rather than that of the whole \CY\ threefold, which results
in a relatively simple determination of the stable spectra.

The derived category can be a technically intimidating subject to
those unfamiliar with the language of homological algebra. In this
paper we attempt to keep the technical details to a minimum. While
this paper was derived as consequence of the results of a previous paper
\cite{AKH:m0}, we hope this paper can be understood to a large extent
without a knowledge of this earlier work.

We should note that the paper \cite{Fiol:suN} is similar to this paper
in many ways. There the BPS spectrum was obtained by using quivers and
$\theta$-stability in the same geometric engineering context that we
use. $\Pi$-stability is a more general picture than $\theta$-stability
so this paper can be thought of as a completion of the work in
\cite{Fiol:suN} to the whole moduli space.

In section \ref{s:mod} we review the basic setup of the geometrical
engineering required to analyze the Seiberg--Witten theory. Section
\ref{s:stab} consists of the analysis of the stable spectrum. This
analysis is performed for three regions of the moduli space. Firstly,
in section \ref{ss:weak}, near the large radius limit. Then we cross
the line of marginal stability into the strongly-coupled regime in
section \ref{ss:strong}, and then, lastly, we push through back into the
weak-coupling region in section \ref{ss:mon}. For this last
computation we are required to go beyond the derived category of
$\P^1$ as we explain in section \ref{ss:vs}.
 

\section{The Model}  \label{s:mod}

\subsection{Engineering the $\SU(2)$}  \label{ss:eng}

In this section we quickly review how to obtain a pure $\SU(2)$
supersymmetric Yang--Mills theory from a type IIA string compactified
on a \CY\ threefold $X$. 

Assume $X$ is a K3 fibration over a base $C\cong\P^1$ with a
section. If $C$ (i.e. the section) is large, then we may expect any
nonabelian gauge symmetry to arise essentially from the type IIA
string compactified on a K3 \cite{me:en3g,KMP:enhg}. One way for this
to occur is the appearance of an A-D-E singularity in the K3
fibres. Another way is that the K3 fibres acquire just the right
volume. These situations are essentially T-dual to each other (see,
for example, \cite{me:lK3} for a review).

In the case that singularities appear in the K3 fibres one may focus
attention near these singularities and analyze this non-compact
region. This is generally the method used in ``geometric engineering''
\cite{KKV:geng}. While we could do that here, for simplicity of
exposition, we will use the case where the K3 fibres acquire a
particular volume to achieve an enhanced gauge symmetry.
We will use the well-known example of \cite{CDFKM:I}, where $X$ is a
degree 8 hypersurface in the resolution of the weighted projective space
$\P^4_{\{2,2,2,1,1\}}$. 

In order to analyze the quantum-corrected moduli space of complexified
K\"ahler forms on $X$ we require an analysis of the mirror $Y$.  $Y$
is given as a $(\Z_4)^3$ quotient of the same hypersurface with
defining equation
\begin{equation}
  a_0z_1z_2z_3z_4z_5 + a_1z_1^4 + a_2z_2^4 + a_3z_3^4 + a_4z_4^8
  + a_5z_5^8 + a_6z_4^4z_4^4.  \label{eq:def1}
\end{equation}
The ``algebraic'' coordinates on the moduli space of complex
structures of $Y$ are then given by
\begin{equation}
  x=\frac{a_1a_2a_3a_6}{a_0^4},\quad y=\frac{a_4a_5}{a_6^2}.
        \label{eq:algco}
\end{equation}
Now use the mirror map to relate these coordinates to the complexified
K\"ahler form on $X$. Let $(B+iJ)_x$ represent the component proportional
to the area of a curve in the K3 fibre, and let $(B+iJ)_y$ represent
the component giving the area of the base $C$. Then
\begin{equation}
\begin{split}
  (B+iJ)_x &= \frac1{2\pi i}\log x + O(x,y)\\
  (B+iJ)_y &= \frac1{2\pi i}\log y + O(x,y),
\end{split}
\end{equation}
where the higher-order terms are obtained from the Picard--Fuchs
equations in the usual way.

Next we compute the locus of ``bad'' conformal field theories
$\Delta$, which can also be thought of as the locus where $Y$ becomes singular.
The ``primary'' component of $\Delta$ can  be computed as
\begin{equation}
  \Delta_0 = (1-2^8x)^2 - 2^{18}x^2y.
\end{equation}
There is also another component
\begin{equation}
  \Delta_1 = 1-4y,
\end{equation}
with $\Delta=\Delta_0\Delta_1^3$.

The enhanced gauge symmetry will not actually appear
unless one is at the limit where $C$ becomes infinitely large,
i.e. $y=0$. For such a nonperturbative effect we must lie on the
discriminant locus, forcing $x=2^{-8}$.

To probe the quantum effects of $\SU(2)$ Yang--Mills theories we copy
the trick from \cite{KKL:limit} and do a double scaling limit to zoom
in on this point $(x,y)=(2^{-8},0)$. In particular we set
\begin{equation}
  u = \frac{1-2^8x}{2\sqrt{y}},
\end{equation}
and take the limit $y\to0$. $\Delta_0$ now intersects the $u$-plane at
$u=\pm1$. This is the $u$-plane of Seiberg and Witten \cite{SW:I}.

We will now use monodromy to establish which solitons are becoming
massless where. Traditionally this is done using matrices denoting
the monodromy action on {\em charges}. We want to know the monodromy
action on the solitons themselves rather than just the charges, so we
use Fourier--Mukai transforms. In particular we rely heavily on the
results and conjectures of \cite{AKH:m0}.

First set $y=0$ and perform monodromy around $x=2^{-8}$. As described
in the main example of \cite{AKH:m0}, this corresponds to an
``EZ-transform'' \cite{Horj:EZ} associated to the fibration $\pi:X\to
C$. According to \cite{AKH:m0} {\em the D-branes on $X$ which become
massless at this point are generated by $\DC(C)$, the derived category
of $C$}. In particular, the basic objects of interest in $\DC(X)$ are
$\pi^*\mathsf{z}$ for $\mathsf{z}\in\DC(C)$.

This statement is key to this paper. Because of the scaling limit we
are taking, any D-brane which is not massless at $(x,y)=(2^{-8},0)$
has effectively infinite mass and will be ignored. Thus all the
solitons of interest in $\SU(2)$ gauge theory are associated to the
derived category of $C\cong\P^1$. The simplicity of the soliton
spectrum of $\SU(2)$ supersymmetric gauge theory is thus related to
the fact that the derived category of $\P^1$ is easy to analyze. 

One cannot ask for a simpler nontrivial space to analyze as far as the
derived category is concerned.  In fact, one may build all objects in
$\DC(C)$ from two basic objects: $\O_C$ --- the structure sheaf of
$C$, and $\O_p$ --- the skyscraper sheaf of a point $p\in C$. That is,
all objects can be constructed by binding together (in a sense to be
described) and translating these basic ingredients.


\subsection{Identifying the central charges} \label{ss:idZ}

In order to analyze the stability of solitons we need to find the central
charges of the relevant objects in the derived category. These central
charges are determined by  periods, and so we need to solve the Picard--Fuchs
equations. For our Calabi--Yau threefold $X$ 
the relevant Picard--Fuchs differential operators are (with
$\Theta_x= x\partial_x$, and similarly for $\Theta_y$) \cite{CDFKM:I,HKTY:}
\begin{equation}\label{eq:e1}
\begin{split}
&\Theta_x^2(\Theta_x-2\Theta_y)-{4}(4\Theta_x+3)(4\Theta_x+2)(4\Theta_x+1)\,,\\
&\Theta_y^2 -{4}(\Theta_x-2\Theta_y)(\Theta_x-2\Theta_y-1)\,.
\end{split}
\end{equation}

Following the analysis of \cite{KKV:geng,KKL:limit} we introduce the
variables
\begin{equation}
\begin{split}
x_1={\frac {4y}{ \left(2^8 x-1 \right) ^{2}}}\,, \quad
x_2=2^8x-1\,.
\end{split}
\end{equation}
For the field theory limit we scale $x_2\to 0$, while $x_1=u^{-2}$ assumes  
arbitrary values. 

In terms of the new variables, the second Picard--Fuchs operator in
(\ref{eq:e1}) becomes
\begin{multline}\label{eq:e3}
-\ff14 x_{{1}}{x_{{2}}}^{2} \left( x_{{2}}+1 \right) ^{2}\p_2^2
-{x_{{1}}}^{2} \left( 4 x_{{1}}{x_{{2}}}^{2}+
4x_{{1}}x_{{2}}+x_{{1}}-1 \right)  \p_1^2\\
+{x_{{1}}}^{2}x_{{2}} \left( x_{{2}}+1 \right) 
 \left( 2x_{{2}}+1\right)  \p_1\p_2
-\ff12x_{{1}} \left( -2+10x_{{1}}x_{{2}}+
10x_{{1}}{x_{{2}}}^{2}+3x_{{1}} \right)\p_1\,.
\end{multline}

For brevity we focus on the semiclassical regime of the gauge theory,
i.e. ${u}=\infty$, although, as pointed out in \cite{KKV:geng},
the monopole and dyon regimes allow for a similar
treatment. Accordingly, the solutions we seek are 
power series expansions around $(x_1,x_2)=(0,0)$. It turns out that
the Picard--Fuchs
system (\ref{eq:e1}) has four solutions with index $(0,0)$ and two solutions
with index $(0,1/2)$. Instead of listing all the solutions, we focus
on the most interesting one:
\begin{equation}
\sqrt {x_2} \left[ 
1-\ff1{16}\,x_{{1}}-{\ff{15}{1024}}\,{x_{{1}}}^{2}
-{\ff{105}{16384}}\,{x_{{1}}}^{3}+O (x_1^4)\right]
+x_{{2}}^{3/2} \left[ 84+O (x_{{1}}) \right]+
O (x_{{2}}^{5/2})\,.
   \label{eq:ser1}
\end{equation}
Up to a constant, these are the first few terms of the Seiberg-Witten
period $a$, at least to leading order in $x_2$. We specialize our
analysis for this solution as follows. In the $x_2\to 0$
limit the leading terms of (\ref{eq:e3}) are\footnote{ The first
operator in (\ref{eq:e1}) vanishes identically in this limit.}
\begin{equation}
-\ff32x_{{1}} \left(x_{{1}} -1 \right)\p_1
+\ff14x_{{1}}x_{{2}}^{2}\,\p_2 -
\ff14x_{{1}}x_{{2}}^{2}\,\p_2^2 
+\ff1{16}-{x_{{1}}}^{2} \left( x_{{1}} -1 \right)\p_1^2 
+{x_{{1}}}^{2}x_{{2}}\,\p_1\p_2,
\end{equation}
and this results in the relevant part of the solution (\ref{eq:ser1})
satisfying the ODE
\begin{equation}\label{eq:e4}
{x_1}^{2} \left( 1-x_1 \right) {\frac {d^{2}}{d{x_1}^{2}}} +\ff32\,x_1 
\left( 1-x_1 \right) {\frac {d}{dx_1}}+\ff1{16}\,.
\end{equation}
Recalling that $x_1=1/u^2$, we can make a final change of
variables $z=({u}-1)/2$, transforming (\ref{eq:e4}) into
\begin{equation}
  z(1-z)\frac{d^2\Phi}{dz^2}-\frac{\Phi}4=0  \,. 
  \label{eq:SWde}
\end{equation}

The central charge of a D-brane is a function of the D-brane
charge. The D-brane charge is measured by K-theory. The K-group of
$\P^1$ is $\Z^2$, generated by $H^0(C)$ and $H^2(C)$. We may take the
two generators of $\DC(C)$, namely $\O_C$ and $\O_p$, to be the
generators of the K-group. The Chern characters of the sheaves
generate $H^0(C)$ and $H^2(C)$ respectively.
We therefore require only two basic central charges
\begin{equation}
\begin{split}
  a &= Z(\pi^*\O_p)\\
  a_D &= Z(\pi^*\O_C),
\end{split}
\end{equation}
to determine the central charges of all solitons. The notation has
been chosen to coincide with Seiberg and Witten \cite{SW:I}. 
The central charges should be ratios of periods, i.e., ratios of
solutions to (\ref{eq:e1}). In the geometric engineering case, the
``denominator'' period in these ratios is taken to be the one
associated to the 0-brane, which is constant in the scaling
limit. This means that the central charges must satisfy
(\ref{eq:SWde}) {\em up to an additive constant}.
This agrees with Seiberg--Witten theory (see \cite{Lc:SWintro} for a
review and references). 

$a$ and $a_D$ may be determined by monodromy as follows. At $u=\pm1$
the primary component of the discriminant hits the $u$-plane. As
discussed in \cite{AKH:m0}, there is considerable evidence that the
structure sheaf $\O_X=\pi^*\O_C$ corresponds to a massless soliton in
this case. This statement is basepoint dependent however. What we
should really do is choose a basepoint close to, say, $u=1$ and say
that $\O_X$ becomes massless there. Thus $a_D=0$ at $u=1$.

\iffigs
\def\mcurve{\closecurve(
  -100,0,  -90,-35,  -70,-60, -40,-78, -20,-84, 
  0,-86,  20,-84,  40,-78,  70,-60, 90,-35,
  100,0, 90,35,  70,60, 40,78, 20,84,
  0,86,  -20,84,  -40,78,  -70,60, -90,35)}
\def\arw#1#2#3#4{{
    \def\xscale{#3}
    \def\yscale{#3}
    \def\xscaley{-#4}
    \def\yscalex{#4}
    \put(#1,#2){\curve(0,0, 2,0.7, 3,2)}
    \put(#1,#2){\curve(0,0, 2,-0.7, 3,-2)}
}}
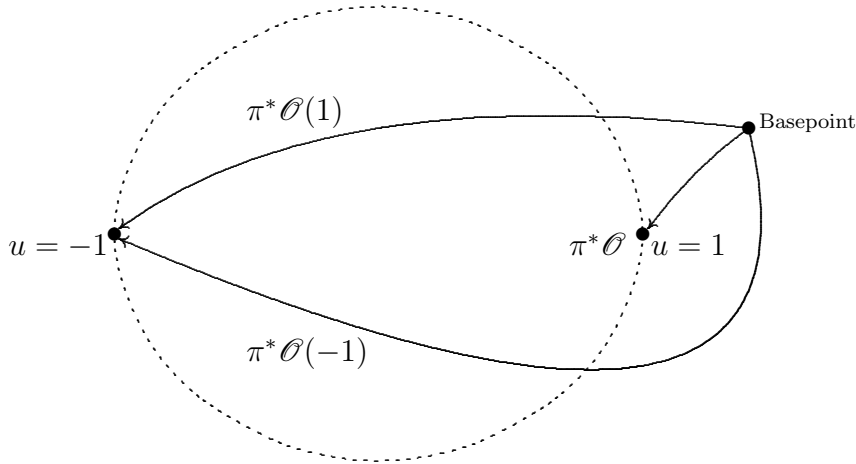
\begin{figure}
\setlength{\unitlength}{1.0pt}
\begin{center}
\begin{picture}(240,200)(-120,-100)
  \curvedashes{3,1}
  \mcurve
  \curvedashes{}
  \put(-100,0){\circle*{5}}
  \put(100,0){\circle*{5}}
  \put(72,-8){\text{$\pi^*\O$}}
  \put(103,-8){\text{$u=1$}}
  \put(-140,-8){\text{$u=-1$}}
  \put(140,40){\circle*{5}}
  \put(144,40){\text{\scriptsize Basepoint}}
  \put(140,40){\curve(0,0, -20,-17, -38,-38)}
     \arw{102}{2}{0.643}{0.766}
  \put(-50,43){\text{$\pi^*\O(1)$}}
  \put(140,40){\curve(0,0, -140,0, -238,-38)}
     \arw{-98}{2}{0.788}{0.616}
  \put(-50,-48){\text{$\pi^*\O(-1)$}}
  \put(140,40){\curve(0,0, -40,-90, -238,-42)}
     \arw{-98}{-2}{0.940}{-0.342}
\end{picture}
  \caption{The massless solitons.}
  \label{f:paths1}
\end{center}
\end{figure}
\fi

Monodromy around $u=-1$ is conjugate, but not isomorphic, to monodromy
around $u=1$. Exactly which D-brane becomes massless depends on the
path from the basepoint to $u=-1$. Let us focus on the ``direct''
paths which do not encircle $u=1$. Then we have essentially two
choices of path which go either side of $u=1$.  As explained in detail
in \cite{AKH:m0}, for example, the Fourier--Mukai transform associated
to monodromy around $u=-1$ may be derived by composing other
Fourier--Mukai transforms. In our case, these rules dictate that
$\pi^*\O_C(\pm1)$ becomes massless at $u=-1$ depending on our choice
of path as shown in figure \ref{f:paths1}. Since
$\ch(\pi^*\O_C(\pm1))=\pi^*(\ch(\O_C)\pm\ch(\O_p))$, this amounts to
$a_D\pm a$ vanishing at $u=-1$. Up to an overall multiplicative
constant, this determines $a$ and $a_D$ uniquely to coincide with
those given in \cite{SW:I}.

This identification with the original work by Seiberg and Witten shows
that the structure sheaf $\O_X=\pi^*\O_C$ corresponds to the
``magnetic monopole'', i.e., the 6-brane wrapping $X$. The other
generator $\pi^*\O_p$ has the right D-brane charge to be associated
with the electrically charged ``W-boson''. In section \ref{ss:weak} we
will see that it really is the W-boson since it is stable.  If $S$ is the K3
fibre over the point $p\in C$, then $\pi^*\O_p$ is the structure sheaf
of $S$ extended by zero over the rest of $X$. That is, $\pi^*\O_p$ is
a 4-brane wrapped around a K3 fibre. This is consistent with the
general assumption (for example in \cite{KMP:enhg}) that the W-bosons
arise from D-branes wrapping the K3 fibres (or curves within the K3
fibres).

Classically we have a whole family of W-bosons since we may move $p$
to be any point in $C$. However, as argued in \cite{KMP:enhg} in this
case, quantum mechanics implies that one should count states by computing
the cohomology of the moduli space. This means that there is really only
one W-boson --- not a whole $\P^1$'s worth.

Note that the above analysis shows that the {\em rank\/} of a sheaf on
$\P^1$ gives the ``magnetic charge'' while the {\em degree\/} of a
sheaf gives the ``electric charge''.


\section{The Stable Spectrum} \label{s:stab}

We are now in a position to establish the spectrum of D-branes at any
point in the $u$-plane. We know that the D-branes in question must be
generated by pull-backs to $X$ of objects in the derived category of
$C$. We also know the central charges of such objects. Armed with
$\Pi$-stability we can solve this problem by determining which objects
are stable. Note that the words ``generated by'' above are a little
loaded. We may need to consider morphisms outside $\DC(C)$ but we
postpone this fact until section \ref{ss:mon}.

\subsection{$\Pi$-stability} \label{ss:Pi}

Objects in the derived category are complexes of sheaves. For any such
complex $\mathsf{A}$ we may define $\mathsf{A}[n]$ simply as the same
complex shifted (or ``translated'') $n$ places left. As explained in
\cite{Doug:DC} a simultaneous shift of all D-branes by $n$ is a gauge
symmetry with an odd $n$ representing an exchange of D-branes with
anti-D-branes. Relative shifts do matter however. An open string
stretched between $\mathsf{A}$ and $\mathsf{B}$ is only the same as an
open string stretched between $\mathsf{A}[n]$ and $\mathsf{B}$ if
$n=0$.

In the derived category picture, open strings between a D-brane
$\mathsf{A}$ and a D-brane $\mathsf{B}$ are given by maps in
$\Hom(\mathsf{A}[n],\mathsf{B})$ for any value of $n$. The maps in
$\Hom(\mathsf{A}[-1],\mathsf{B})$ are of particular interest as giving
these strings a vacuum expectation value forms a ``bound'' state of
$\mathsf{A}$ and $\mathsf{B}$ as explained in \cite{AL:DC}.
Mathematically this new object formed by the bound state is called the
``Cone'' of the corresponding map $f:\mathsf{A}[-1]\to\mathsf{B}$. A
nice explanation for this language is given in \cite{Thom:DCg}.
In derived category language this leads to a
``distinguished triangle'':
\begin{equation}
\xymatrix{
&\Cone(f)\ar[dl]|{[1]}&\\
\mathsf{A}[-1]\ar[rr]^f&&\mathsf{B}.\ar[ul]
}   \label{eq:ABC}
\end{equation}
Any vertex of this triangle is a potential bound state of the other
two vertices. We refer to \cite{AD:Dstab} for more details.  Whether
or not each state really is bound or not depends upon the mass of the
open string forming the map on the opposite edge of the triangle. Only
when this open string is tachyonic do we achieve a bound state. The
``$[1]$'' on the arrow in (\ref{eq:ABC}) means that a left-shift of
one is included in the map. This ``[1]'' may be shuffled around to any
edge of the triangle if the objects at the vertices are shifted accordingly.
Such shuffles will occur frequently in the computations below.

The rules of computing the masses of these open strings arise from
$\Pi$-stability as follows \cite{DFR:stab,Doug:DC,AD:Dstab}. Each
stable D-brane is given a ``grade'' $\varphi\in\R$ which varies
continuously over the moduli space. It is not single-valued over the
moduli space however. It is defined mod 2 by the central charge
\begin{equation}
  \varphi(\mathsf{A}) = -\frac1\pi\arg(Z(\mathsf{A})) \pmod2,
    \label{eq:phidef1}
\end{equation}
and $\varphi(\mathsf{A}[n])=\varphi(\mathsf{A})+n$.
To complete the definition of $\varphi$ one can declare that the grade
of a coherent sheaf lies in some fixed range (say, between $-2$ and 0)
near the large radius limit \cite{Doug:DC,Doug:S01,AD:Dstab}.

The mass of an open string stretched from $\mathsf{A}[-1]$ to $\mathsf{B}$
is then proportional to $\varphi(\mathsf{B})-\varphi(\mathsf{A})$
as explained in \cite{Doug:DC}. In this way we may determine the
regions of stability for all potential D-branes.

Points of marginal stability in the $u$-plane can appear wherever
$\varphi(\mathsf{B})-\varphi(\mathsf{A})=0$. Given (\ref{eq:phidef1})
this means that the ratio of the central charges of $\mathsf{A}$ and 
$\mathsf{B}$ is real. For pure $\SU(2)$ there is only a two-dimensional
lattice of possible D-brane charges, i.e., all central charges are
linear combinations of $a$ and $a_D$. Thus the only possible points of
marginal stability occur when $a/a_D$ is real. As is well-known
\cite{Fzdn:msline,AFS:msline}, the set of such points forms a
near-ellipse passing through $u=\pm1$. We show this as a dotted line
in figure \ref{f:paths1} and refer to it as the ``ms-line''.


\subsection{Weak coupling region} \label{ss:weak}

Let us begin ``outside'' the ellipse of marginal stability. Since this
is the only ms-line of concern, we may move to very
large $|u|$ without effecting the stable spectrum. In terms of $X$,
this allows us to move to the large radius limit of $X$ where all
$\alpha'$-corrections are very small.

In the large radius limit we expect the classical analysis of D-branes
such as in \cite{BBS:5b,OOY:Dm} to hold true and B-type D-branes
should correspond to stable holomorphic vector bundles over complex
analytic subspaces of $X$. In other words, coherent sheaves that are
locally free over some complex subspace of $X$ (possibly all of $X$)
and zero elsewhere. Our interest is in sheaves of particular type:
those  that can
be written as the pullback of  a sheaf on $C$ described by the
previous sentence. As explained
earlier, only these sheaves will correspond to solitons associated to
the $\SU(2)$ theory in question.

The classification of such sheaves on $\P^1$ is very easy. First we
have $\O_p$, the skyscraper sheaf of a point. This is classically
stable since it has no subsheaves. This pulls back to $X$
to give the structure sheaf of a K3 fibre. As seen in section
\ref{ss:idZ}, this stable soliton is identified with the W-boson.

The only remaining possibilities are stable vector bundles over
$\P^1$. Thanks to a theorem of Grothendieck \cite{Grot:P1} all vector
bundles of rank $r$ over $C\cong\P^1$ are isomorphic to a direct sum of line
bundles
\begin{equation}
  \cF \cong \O_C(s_1) \oplus \O_C(s_2) \oplus\ldots\oplus \O_C(s_r),
\end{equation}
for integers $s_1,s_2,\ldots ,s_r$.  Such bundles of rank one are
stable. To show this we could use classical $\mu$-stability which is
related to $\Pi$-stability at large radius
\cite{DFR:stab,AD:Dstab}. We choose to employ $\Pi$-stability directly
to keep a more consistent language throughout this paper.
Near the large radius limit we use \cite{MM:K,FW:D}
\begin{equation}
  \varphi(\cF)=-\frac1{\pi}\arg\int_X
  \exp(B+iJ)\ch(\cF)\sqrt{\td(X)}+\ldots,  \label{eq:LRL}
\end{equation}
and fix $-2<\varphi(\cF)\leq0$. It follows that
$\varphi(\pi^*\O_p)=-1+O(\epsilon)$ and
$\varphi(\pi^*\O_C(s))=-\ff32+s\epsilon+\rho(B)\epsilon+O(\epsilon^2)$ where
$\epsilon$ is a some small positive real number which goes to zero in
the large radius limit, and $\rho(B)$ is a $B$ dependent constant.
Note that this need not agree with the
corresponding values computed by using the $a$ and $a_D$ basis of
section \ref{ss:idZ} since the central charge has been left undefined
up to an overall constant. What is true however, is that the relative {\em
differences\/} between grades determined by $a$ and $a_D$ must agree
with this large radius limit.

In order to destabilize $\O_C(s)$ we need to build a distinguished
triangle with $\O_C(s)$ at one corner. This requires a map
$g:\cE\to\O_C(s)$ for some stable sheaf $\cE$. No map exists for
$\cE=\O_p$. Suppose $\cE=\O_C(r)$ for some $r$. A map $g$ only exists if
$r\leq s$. The case of $r=s$ is trivial as this corresponds to
$\O_C(s)$ decaying into itself plus nothing! We may therefore assume
$r<s$. As far as sheaves are concerned, distinguished triangles amount
to short exact sequences completed into a triangle by adding the
arrow with a ``[1]'' in it. This leads to a triangle
\begin{equation}
\xymatrix@C=5mm{
&\O_p^{\oplus(s-r)}\ar[dl]|{[1]}&\\
\O_C(r)\ar[rr]^g&&\O_C(s),\ar[ul]
}   \label{eq:rk1}
\end{equation}
where, by a slight abuse of notation, $\O_p^{\oplus(s-r)}$ is the
skyscraper sheaf of $s-r$ points. But 
\begin{equation}
\varphi(\pi^*\O_C(r))-\varphi(\pi^*\O_p^{\oplus(s-r)})
  = -\ff12+O(\epsilon),
\end{equation}
which is less than zero showing stability. Thus rank one bundles are
stable against decay into other rank one bundles or skyscraper
sheaves. We will soon see that higher rank bundles are unstable so we
do not need to consider them as possible decay products of rank one
bundles.

But wait! The reason higher rank bundles are unstable is because they
will be shown to decay into rank one bundles. Such analysis will
implicitly assume that the rank one bundles are indeed stable, so we
have a circular argument. To be
rigorous we should turn to $\mu$-stability. Here a bundle can only be
unstable with respect to a subbundle. Clearly a rank one bundle cannot
have a higher rank subbundle and so we immediately rule out such
decays. What is at work here is that sheaves form
an {\em abelian\/} category and allow the notion of a subobject. The
derived category has no notion of subobject and so $\Pi$-stability is
always plagued by such circularity. The importance of abelian
categories to establish a set of stable solitons at least at some
basepoint in the moduli space was emphasized by Douglas \cite{Doug:DC}.

We now consider the stability of rank 2 bundles. The bundle
$\O_C(s)\oplus\O_C(t)$ fits into the triangle
\begin{equation}
\xymatrix@C=1mm{
&\O_C(s)\oplus\O_C(t)\ar[dl]&\\
\O_C(s)\ar[rr]|(0.6){[1]}^f&&\O_C(t),\ar[ul]
}   \label{eq:rk2}
\end{equation}
where $f\in\Hom(\O_C(s),\O_C(t)[1])=\Ext^1(\O_C(s),\O_C(t))$.
The direct sum $\O_C(s)\oplus\O_C(t)$ occurs when the map $f$ is the
zero map. If $f$ is any other map, it will deform the direct sum into
some other extension of $\O_C(s)$ by $\O_C(t)$ according to the usual
theory of extensions \cite{HS:hom}. But $\Ext^1(\O_C(s),\O_C(t))$ has
dimension $s-t-1$ if $s\geq t+2$ and is trivial otherwise. Now
\begin{equation}
  \varphi(\O_C(t))-\varphi(\O_C(s)) = \epsilon(t-s)+O(\epsilon^2).
    \label{eq:stbind}
\end{equation}
This implies that if $s<t+2$ then we have no open string along the
bottom edge of (\ref{eq:rk2}) and the direct sum
$\O_C(s)\oplus\O_C(t)$ remains as it is, at least according to this
triangle. If, on the other hand, $s\geq t+2$, then, according to
(\ref{eq:stbind}), we have a tachyon along the bottom edge of the
triangle (\ref{eq:rk2}). One might therefore be tempted to say that
the top vertex of the triangle is stable, but we need to think
carefully about this.

The fact that the open string given by $f$ is tachyonic means that it
wishes to acquire a nonzero expectation value. That is, we need to
``turn on'' $f\in \Ext^1(\O_C(s),\O_C(t))$ to make a nontrivial
extension. This will deform the direct sum $\O_C(s)\oplus\O_C(t)$ into
a rank 2 coherent sheaf, which turns out to be another rank 2 bundle
$\cF$ \cite{Friedm:algsurf}, which is another extension of $\O_C(s)$ by
$\O_C(t)$. But by Grothendieck's theorem, all rank 2 bundles are of
the form $\cF\cong\O_C(s')\oplus\O_C(t')$, for some integers $s'$ and
$t'$. That is, we have a nontrivial extension
\begin{equation}
  0 \to \O_C(t) \to \O_C(s')\oplus\O_C(t') \labto{h} \O_C(s)\to0.
  \label{eq:extG}
\end{equation}
By looking at degrees, it
is clear that $s+t=s'+t'$, while surjectivity of the map $h$ in
(\ref{eq:extG}) forces  $t<t'\leq s'<s$.
Now if $s'\geq t'+2$ then we may repeat the process since we will
still have a tachyon. It is clear then that we only achieve stability
when $s'-t'$ equals 0 or 1. 

Clearly the r\^oles of $s$ and $t$ can be exchanged in this
discussion so we arrive at the conclusion that {\em the rank two bundle
$\O(s)\oplus\O(t)$ will ``decay'' into $\O(s')\oplus\O(t')$ for
$s'+t'=s+t$ and $|s'-t'|\leq1$.}

So suppose we have a rank two bundle $\O(s)\oplus\O(t)$ with
$|s-t|\leq1$. This is composed of the two ``constituents'' $\O(s)$ and
$\O(t)$. Since $\Ext^1(\O_C(s),\O_C(t))=\Ext^1(\O_C(t),\O_C(s))=0$,
there are no open strings to either stabilize or destabilize this
combination. We claim therefore that this state is semistable and the
two constituents are free to drift apart.\footnote{We do not consider
the possibility of marginal bound states here.}  In this sense {\em
no\/} rank 2 bundles form stable bound states.

\iffigs
\begin{figure}
  \centerline{\epsfxsize=9cm\epsfbox{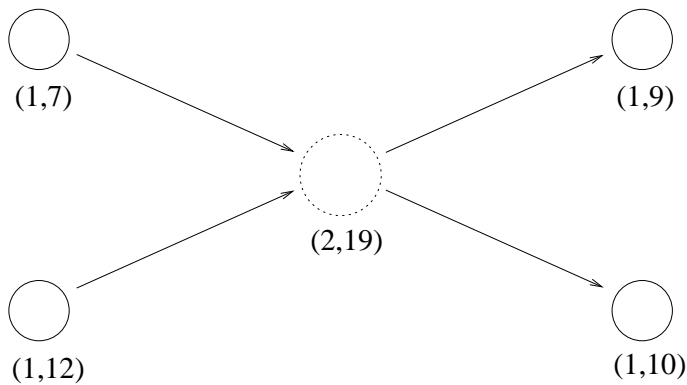}}
  \caption{Dyon Scattering.}
  \label{f:dy}
\end{figure}
\fi

We show the resulting dyon scattering process in a typical case in
figure~\ref{f:dy}. Following the notation of \cite{SW:I}, a sheaf
$\O_C(m)$ will have dyon charge $(n_m,n_e)=(1,m)$. Thus two dyons of charge
$(1,7)$ and $(1,12)$ will have a tachyonic string between them and
will therefore be ``attracted'' to form a bound state. The resulting
dyon will have charge $(2,19)$. This latter state is composed of dyons
of charge $(1,9)$ and $(1,10)$ which do not interact by open strings
and are free to fly apart.

Clearly we may generalize this result to higher rank. Any higher rank
bundle will decay into a set of line bundles whose degrees differ by
at most one. This uniquely identifies the decay products of any bundle.

Any dyon of charge $(r,m)$ for $r>1$ is therefore unstable to decay
into dyons with $r=1$. By a similar reasoning there are no stable
states of charge $(0,m)$ for $m>1$. These would have to correspond to
a bound state of skyscrapers $\O_p$ --- but clearly these skyscrapers
all have the same grade $\varphi$ and so we could never develop
tachyons between them.

The only stable states therefore are $\O_C(m)$ and $\O_p$ which have
dyon charge $(1,m)$ and $(0,1)$ respectively. As always, we also have all the
translates given by $\O_C(m)[n]$ and $\O_p[n]$ for any $n$ which
includes the corresponding anti-particles. This spectrum is in
agreement with known results \cite{SW:I,BF:su2}.


\subsection{Strong coupling region} \label{ss:strong}

Now that we have established a stable set of D-branes near large radius
limit, we may venture to cross the ms-line and see what
$\Pi$-stability tells us. This may be determined purely from the
properties of D-branes from the last section, i.e., D-branes stable
at large radius.

One might fear that, as we cross the ms-line, a new stable soliton
appears that is a bound state of formerly {\em unstable\/}
solitons. Fortunately, thanks to an analysis of the octahedral axiom 
in \cite{AD:Dstab}, this need not be considered. The formerly
unstable solitons must themselves have been some unbound state of
stable solitons which may instead be used as the constituent particles
for the desired new soliton as we cross the ms-line. 

Naturally, what happens as we cross the line of marginal stability
depends on the path taken. Let us first assume that we cross the {\em
upper\/} half of the line of marginal stability by taking the upper
path of figure \ref{f:paths1}.

Integrating the differential equation (\ref{eq:SWde}) we find that
this portion of the ms-line corresponds to $a_D/a$ a real
number with $a_D/a= 0$ at $u=0$ continuously decreasing to $a_D/a=-1$
at $u=-1$. 

We now try to define $\arg(a_D/a)$ as a real number varying
continuously in the $u$-plane in order to obtain the $\varphi$'s of
our D-branes. At large radius limit we may use (\ref{eq:LRL}) to
establish that $\arg(a_D/a)=\pi/2$, fixing the $2n\pi$ ambiguity. As we
approach the ms-line this number increases steadily
to $\pi$.  That is, slightly above this line of marginal stability
$\arg(a_D/a)$ is a little less than $\pi$. As we move past the line,
$\arg(a_D/a)$ passes through $\pi$. To define a continuous function,
consistent with the grading, we say that $\arg(a_D/a)$ is
a  little bit greater than $\pi$.

In order to establish the spectrum of stable D-branes as the
ms-line is crossed we need to consider all possible distinguished
triangles with vertices corresponding to the stable D-branes found in
the last section. All of our objects are of the form $\pi^*\mathsf{z}$
for some $\mathsf{z}\in\DC(C)$. Unfortunately it is not true that all
morphisms $\pi^*\mathsf{x}\to\pi^*\mathsf{y}$ in the derived category
of $X$ can be obtained by pulling back a morphism
$\mathsf{x}\to\mathsf{y}$ as we discuss in section \ref{ss:vs}. For this
section, however, it turns out that 
we can get away with making this assumption and we
need only consider distinguished triangles in the derived category of
$\P^1$.

In fact we only need to consider a single distinguished triangle to
obtain the result. Namely
\begin{equation}
\xymatrix{
&\O_p\ar[dl]|{[1]}_(0.4)h&\\
\O_C(m)\ar[rr]^f&&\O_C(m+1),\ar[ul]_g
}   \label{eq:tri}
\end{equation}
with the ``[1]'' shuffled around as needed. (Actually all distinguished
triangles in the derived category of $\P^1$ may be written in terms of
(\ref{eq:tri}).)  We now analyze the masses of the open strings
corresponding to the morphisms $f$, $g$, and $h$.

We begin with $g$. One obtains immediately
\begin{equation}
\begin{split}
  \varphi(\pi^*\O_p[-1])-\varphi(\pi^*\O_C(m+1)) &=
     -1-\frac1{\pi}\arg(a)+\frac1{\pi}\arg((m+1)a + a_D)\\
  &= -1+\frac1{\pi}\arg\left(\frac{a_D}a+m+1\right).
\end{split}
\end{equation}
Now, since $-1<a_D/a<0$, if $m\geq0$ we have a difference in grading
of around $-1$. Thus the open string is tachyonic and $\O_C(m)$ will
not decay into $\O_C(m+1)$ and $\O_p[-1]$. If on the other hand $m<0$,
then the difference in grading will increase through 0 as the ms-line
is crossed. Then $\O_C(m)$ will decay into $\O_C(m+1)$ and $\O_p[-1]$.

Thus, for example, $\O_C(-2)$ will decay into $\O_C(-1)$ and
$\O_p[-1]$. But $\O_C(-1)$ itself will also decay into $\O_C$ and
$\O_p[-1]$. Therefore $\O_C(-2)$ will actually decay into $\O_C$ and
two copies of $\O_p[-1]$, assuming that these final products are
stable. Anyway, at this point we have shown that all states $\O_C(m)$
for $m<0$ will decay.

Now consider $h$. One obtains
\begin{equation}
  \varphi(\pi^*\O_C(m))-\varphi(\pi^*\O_p)
  = -\frac1{\pi}\arg\left(\frac{a_D}a+m\right).
\end{equation}
Now if $m\leq0$, this difference is close to $-1$ and so we have
stability. However, if $m>0$ then this difference will increase
through 0 as we cross the ms-line. Thus
$\O_C(m+1)$ will then decay into $\O_C(m)$ and $\O_p$. In other words,
all states $\O_C(m)$ for $m>1$ will decay by this route.

At this point we see that $\O_C(m)$ is only stable if $m=0$ or 1. Thus
to check for stability of $\O_p$ using the map $f$ in (\ref{eq:tri})
we need only consider $m=0$. Then
\begin{equation}
  \varphi(\pi^*\O_C(1))-\varphi(\pi^*\O_C[1])
  = -1-\frac1{\pi}\arg\left(\frac{a}{a_D}+1\right).
\end{equation}
As we cross the ms-line this difference increases
through zero. That is, $\O_p$ decays into $\O_C(1)$ and $\O_C[1]$.

The final result is that the only stable solitons after we cross the
ms-line are $\O_C$ and $\O_C(1)$ (and their translates corresponding to
anti-solitons).  Thus, in the example above, $\O_C(-2)$ would actually
decay into three copies of $\O_C$ and two copies of $\O_C(1)[-1]$.
Note that we have only dyon charges $(1,0)$ and $(1,1)$ in agreement 
again with known results \cite{SW:I,BF:su2}.

We may repeat the above analysis for the {\em lower\/} half of the
ms-line by taking the lower path of figure \ref{f:paths1}. The only
difference in the analysis is that now $0< a_D/a<1$ when we hit the
ms-line. The result now is that the only stable solitons consist of
$\O_C$ and $\O_C(-1)$. This is in agreement with the fact that monodromy 
around the monopole point $u=1$ transforms $\O_C(1)$ into $\O_C(-1)[1]$.


\subsection{Returning to weak coupling} \label{ss:mon}

Suppose now we continue our journey and pass through the other wall of
the region bounded by the ms-line and return to the weakly coupled
region as shown by the path in figure \ref{f:paths2}. One should
return to a set of stable solitons physically equivalent to those
found in section \ref{ss:weak}, but they may not actually be the same
elements of the derived category. That is, we undergo some monodromy
in the spectrum.

Monodromy has been analyzed extensively using Fourier--Mukai
transforms (see \cite{AKH:m0} and references therein) but, rather than
use this technology, we will just continue the above analysis. 

\iffigs
\begin{figure}
\setlength{\unitlength}{1.0pt}
\begin{center}
\begin{picture}(240,200)(-120,-100)
  \curvedashes{3,1}
  \mcurve
  \curvedashes{}
  \put(-100,0){\circle*{5}}
  \put(100,0){\circle*{5}}
  \put(103,-8){\text{$u=1$}}
  \put(-140,-8){\text{$u=-1$}}
  \put(140,40){\circle*{5}}
  \put(144,40){\text{\scriptsize Basepoint}}
  \put(0,0){\curve(140,40, 50,70, 0,-100)}
    \arw{0}{-100}{0.174}{0.985}
  \put(50,90){\text{\scriptsize $\pi^*\O_p$, $\pi^*\O_C(m)$}}
  \put(-50,0){\text{\scriptsize $\pi^*\O_C$, $\pi^*\O_C(1)$}}
  \put(2,-103){\text{\scriptsize $\cX$, $\cY_m$}}
\end{picture}
  \caption{The stable solitons as we follow a path.}
  \label{f:paths2}
\end{center}
\end{figure}
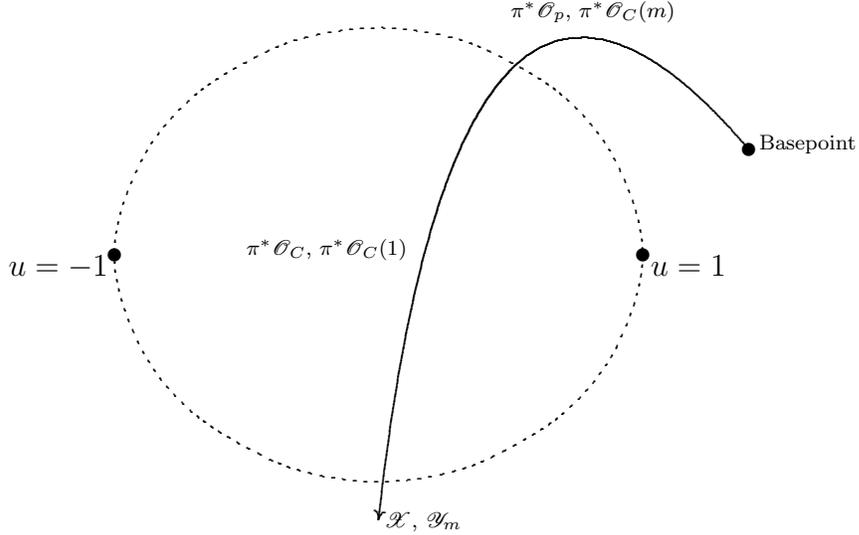
\fi

All the distinguished triangles in $\DC(C)$ are basically obtained
from (\ref{eq:tri}). Unfortunately, if we use the analysis of section
\ref{ss:strong}, then nothing happens to the spectrum as we cross back
into the weakly-coupled region. In particular, the open string
corresponding to a map $\O_C\to\O_C(1)$ has mass squared proportional to
$\varphi(\pi^*\O_C(1))-\varphi(\pi^*\O_C[-1])$, which simply gets
larger and larger as we follow the path in figure \ref{f:paths2}. The
bound state, which gives $\O_p$, becomes unstable as we enter the
strong coupled region and is even more unstable as we leave again.

The problem is that we now need to worry about open strings arising
from morphisms that cannot be pulled back from $\DC(C)$. In
particular, for $m\leq1$, one can show that
\begin{equation}
  H^3(X,\pi^*\O_C(m)) = \C^{1-m},
\end{equation}
where $H^3(X,\pi^*\O_C(m))\cong\Ext^3(\pi^*\O_C(-m),\pi^*\O_C)\cong
\Hom(\pi^*\O_C(-m),\pi^*\O_C[3])$. That is, we have a 2-dimensional
Hilbert space of open strings from $\pi^*\O_C(1)[-1]$ to
$\pi^*\O_C[2]$. Now, analysis of the periods shows that
$\varphi(\pi^*\O_C[2])-\varphi(\pi^*\O_C(1))$ decreases as we follow
the path and falls below zero as we hit the lower portion of the
ms-line in figure \ref{f:paths2}. Thus we acquire new bound states of
$\pi^*\O_C(1)$ and $\pi^*\O_C[2]$ which we call $\cX$. Classically we
have a $\P^1$'s worth of these states $\cX$ just like the old
$\O_p$'s. This is very reminiscent of the 0-branes living on flopped
$\P^1$'s analyzed in \cite{Brig:flop,me:point}. $\cX$
sits in the following distinguished triangle:
\begin{equation}
\xymatrix{
&\cX\ar[dl]|{[1]}&\\
\pi^*\O_C(1)[-1]\ar[rr]&&\pi^*\O_C[2].\ar[ul]
}   \label{eq:triX}
\end{equation}
Consider this to be a special case of the following distinguished triangle
\begin{equation}
\xymatrix{
&\cX\ar[dl]|{[1]}&\\
\cY_m\ar[rr]&&\cY_{m+1},\ar[ul]
}   \label{eq:triXn}
\end{equation}
where $\cY_0=\pi^*\O_C[2]$ and $\cY_{-1}=\pi^*\O_C(1)[-1]$. This may
be used to  define $\cY_{m+1}$ inductively as a bound state of
$\cY_m$ and $\cX$, or $\cY_{m-1}$ as a bound state of $\cY_m$ and
$\cX[-1]$. This defines (at least up to isomorphism) objects $\cY_m$
in $\DC(X)$ for $m=-\infty\ldots\infty$.

Working through the gradings using the rules of \cite{AD:Dstab} it is
not hard to show that {\em all\/} of the $\cY_m$'s become stable as we
pass below the ms-line back into the weak-coupling regime. There are
no other homomorphisms between these stable objects of the right
relative grades to produce any new stable states. Thus the stable
spectrum consists of $\cX$ and $\cY_m$ for all $m$.

Of course these states should correspond to the original set of stable
states under monodromy around $u=1$, i.e., the primary component of
the discriminant. One can use the transforms discussed in
\cite{ST:braid} to show that this is indeed the case. It is easy to compute
the dyon charges of $\cX$ and $\cY_m$ as $(2,1)$ and $(1+2m,m)$
respectively which is in agreement with the known monodromy matrices
\cite{SW:I}.

It is worth dwelling on a couple of points raised in the analysis of
this section. First let us consider the ``cascade'' of new states
coming from (\ref{eq:triXn}) as we crossed the ms-line. 

The rules of computing grades analyzed in \cite{AD:Dstab} said that
when a new bound state is formed, the grade of the new state is
identical to the grades of the constituent states (shifted by 1 as
necessary). Thus if $\mathsf{A}$ can bind to $\mathsf{B}$ then the
grades would imply $\mathsf{A}$ can bind to the resulting bound
state. Thus $\mathsf{A}$ appears able to bind an arbitrary number of
times to $\mathsf{B}$. However, this can only happen if open strings
exist to produce the necessary bindings. In particular, just because
there is a morphism from $\mathsf{A}$ to $\mathsf{B}$, there might not
be a morphism from $\mathsf{A}$ to the bound state of $\mathsf{A}$ and
$\mathsf{B}$. Then only one copy of $\mathsf{A}$ would be able to bind
to $\mathsf{B}$. The relationship between the dimensions of the
various Hom groups can be computed using long exact sequences
associated to the triangles. The magic that happens in
(\ref{eq:triXn}) is that these long exact sequences work out just
right to make $\dim\Hom(\cX,\cY_m[1])$ and $\dim\Hom(\cY_{m+1},\cX)$
independent of $m$. It is this property that allows us to add $\cX$'s
indefinitely to the D-branes $\cY_m$.

The other point is that the original set of states at weak coupling
all look nice in the sense that they have clear interpretations in
terms of D-branes wrapped on cycles in $X$, whereas after monodromy we
have more exotic objects in the derived category that cannot (except
for $\cY_{-1}$ and $\cY_0$) be written as sheaves. Thus if we sit in the
strong coupling region there seems to be an asymmetry between moving
up or moving down into the weak coupling regime.

However, if we view things purely in the derived category they look
more symmetric. When we sit in the strong-coupling region we have only
two stable D-branes which we denote $\mathsf{A}$ and $\mathsf{B}$. As
we move across the ms-line into the weak coupling regime these two
solitons can form a bound state $\mathsf{W}$ which plays the r\^ole of
the W-boson. Crossing the ms-line upwards this bound state is formed
by condensing a tachyon given by the open string associated to the map
$\mathsf{A}\to\mathsf{B}$. Crossing the ms-line downwards is exactly
the same except this time $\mathsf{W}$ is formed from the open string
$\mathsf{B}\to\mathsf{A}[3]$. This is essentially Serre duality at
work and is very similar to the examples studied in \cite{AD:Dstab}
for finding exotic solitons on the quintic threefold. It is also
entertaining to note that the ``3'' appearing here shows that the
intrinsic structure of the solitons in Seiberg--Witten theory ``knows''
that it should be associated with a \CY\ of dimension three!

Once we have crossed into the weak coupling regime, $\mathsf{W}$ may
bind to an arbitrary number of copies of $\mathsf{A}$ or $\mathsf{B}$ to
form the remaining stable solitons. This description is valid
independent of any basepoint chosen. The basepoint is only required to
explicitly describe the solitons as complexes of sheaves.


\subsection{$\DC(C)$ versus $\DC(X)$} \label{ss:vs}

Finally let us note again that in sections \ref{ss:weak} and
\ref{ss:strong} we were able to couch our description of the D-branes
purely in terms of the derived category of the base $C\cong\P^1$ of
the K3-fibration. This explains the relative simplicity of the
analysis of solitons in Seiberg--Witten theory. The complete
description, in terms of $\DC(X)$, may add more morphisms than are
seen in $\DC(C)$ but these extra open strings play no r\^ole in
tachyon condensation near large radius. Once we venture ``too far''
however by looking at monodromy in section \ref{ss:mon} we need the
full derived category of $X$ to obtain the correct soliton spectrum.

This may be understood schematically as follows. For two objects
$\mathsf{a}$ and $\mathsf{b}$ in $\DC(C)$ we have
\begin{equation}
  \Hom_{\DC(X)}(\pi^*\mathsf{a},\pi^*\mathsf{b})
    = \Hom_{\DC(C)}(\mathsf{a},\pi_*\pi^*\mathsf{b}),
\end{equation}
and thus the difference between the open strings in $\DC(C)$ and the
open strings in $\DC(X)$ essentially corresponds to the difference
between $\mathsf{b}$ and $\pi_*\pi^*\mathsf{b}$.

At the large radius limit we assume that all B-type D-branes
correspond to sheaves, i.e., complexes in $\DC(X)$ which have only one
nonzero entry. We also confine the relative grades to small range
$-2<\varphi\leq0$. Now $\pi_*\pi^*\mathsf{b}$ is usually a longer
complex than the single entry $\mathsf{b}$ but in order to ``see'' the
new terms in the complex in terms of tachyon condensation one needs
larger relative grades. Thus one needs to travel a large distance from
the large radius limit to detect the new terms in
$\pi_*\pi^*\mathsf{b}$. This is precisely what happened in section
\ref{ss:mon}. The purely $\DC(C)$ morphisms gave us ``enough'' information
to cover the whole moduli space once. It was only
monodromy issues that showed $\DC(C)$'s inadequacies.

It would be interesting to see if the pure $\DC(C)$ picture of the
solitons in Seiberg--Witten theory persists for more complicated
examples of $N=2$ theories beyond the pure $\SU(2)$ theory considered
here. Investigations are currently underway in the case of $\SU(3)$
and $\SU(2)$ with fundamental matter. We hope to report on these
in a future publication.


\section*{Acknowledgments}

It is a pleasure to thank P.~Horja, K.~Narayan, R.~Plesser, H. Schenck
and M. Stillman for useful
conversations.  P.S.A.\ is supported in part by NSF grant DMS-0074072
and by a research fellowship from the Alfred P.~Sloan
Foundation. R.L.K. is partly supported by NSF grants DMS-9983320 and
DMS-0074072.


\begin{thebibliography}{10}

\bibitem{SW:I}
N.~Seiberg and E.~Witten,
\newblock {\em Electric - Magnetic Duality, Monopole Condensation, and
  Confinement in N=2 Supersymmetric Yang-Mills Theory},
\newblock Nucl. Phys. {\bf B426} (1994) 19--52, hep-th/9407087,
\newblock (erratum-ibid. {\bf B430} (1994) 485-486).

\bibitem{BF:su2}
F.~Ferrari and A.~Bilal,
\newblock {\em The Strong-Coupling Spectrum of the Seiberg--Witten Theory},
\newblock Nucl. Phys. {\bf B469} (1996) 387--402, hep-th/9602082.

\bibitem{AKM:lcy}
P.~S. Aspinwall, S.~Katz, and D.~R. Morrison,
\newblock {\em Lie Groups, Calabi--Yau Threefolds and F-Theory},
\newblock Adv. Theor. Math. Phys. {\bf 4} (2000) 95--126, hep-th/0002012.

\bibitem{KKL:limit}
S.~Kachru et~al.,
\newblock {\em Nonperturbative Results on the Point Particle Limit of N=2
  Heterotic String Compactifications},
\newblock Nucl. Phys. {\bf B459} (1996) 537--558, hep-th/9508155.

\bibitem{KKV:geng}
S.~Katz, A.~Klemm, and C.~Vafa,
\newblock {\em Geometric Engineering of Quantum Field Theories},
\newblock Nucl. Phys. {\bf B497} (1997) 173--195, hep-th/9609239.

\bibitem{KLM:hSW}
A.~Klemm et~al.,
\newblock {\em Self-Dual Strings and N=2 Supersymmetric Field Theory},
\newblock Nucl. Phys. {\bf B477} (1996) 746--766, hep-th/9604034.

\bibitem{Lc:SWintro}
W.~Lerche,
\newblock {\em Introduction to Seiberg--Witten Theory and its Stringy Origin},
\newblock Nucl. Phys. Proc. Suppl. {\bf 55B} (1997) 83--117, hep-th/9611190.

\bibitem{RSVV:stab}
A.~Ritz, M.~A. Shifman, A.~I. Vainshtein, and M.~B. Voloshin,
\newblock {\em Marginal Stability and the Metamorphosis of BPS States},
\newblock Phys. Rev. {\bf D63} (2001) 065018, hep-th/0006028.

\bibitem{AN:webs}
P.~C. Argyres and K.~Narayan,
\newblock {\em String Webs from Field Theory},
\newblock JHEP {\bf 03} (2001) 047, hep-th/0101114.

\bibitem{Dnf:Dstab}
F.~Denef,
\newblock {\em Supergravity Flows and D-brane Stability},
\newblock JHEP {\bf 08} (2000) 050, hep-th/0005049.

\bibitem{DGR:Dquin}
F.~Denef, B.~Greene, and M.~Raugas,
\newblock {\em Split Attractor Flows and the Spectrum of BPS D-branes on the
  Quintic},
\newblock JHEP {\bf 05} (2001) 012, hep-th/0101135.

\bibitem{Kon:mir}
M.~Kontsevich,
\newblock {\em Homological Algebra of Mirror Symmetry},
\newblock in ``Proceedings of the International Congress of Mathematicians'',
  pages 120--139, Birkh{\"a}user, 1995,
\newblock alg-geom/9411018.

\bibitem{Doug:DC}
M.~R. Douglas,
\newblock {\em D-Branes, Categories and $N$=1 Supersymmetry},
\newblock J. Math. Phys. {\bf 42} (2001) 2818--2843, hep-th/0011017.

\bibitem{Laz:DC}
C.~I. Lazaroiu,
\newblock {\em Unitarity, D-Brane Dynamics and D-brane Categories},
\newblock JHEP {\bf 12} (2001) 031, hep-th/0102183.

\bibitem{AL:DC}
P.~S. Aspinwall and A.~E. Lawrence,
\newblock {\em Derived Categories and Zero-Brane Stability},
\newblock JHEP {\bf 08} (2001) 004, hep-th/0104147.

\bibitem{Dia:DC}
D.-E. Diaconescu,
\newblock {\em Enhanced D-brane Categories from String Field Theory},
\newblock JHEP {\bf 06} (2001) 016, hep-th/0104200.

\bibitem{DFR:stab}
M.~R. Douglas, B.~Fiol, and C.~R{\"o}melsberger,
\newblock {\em Stability and BPS Branes},
\newblock hep-th/0002037.

\bibitem{DFR:orbifold}
M.~R. Douglas, B.~Fiol, and C.~Romelsberger,
\newblock {\em The Spectrum of BPS Branes on a Noncompact Calabi-Yau},
\newblock hep-th/0003263.

\bibitem{Doug:Dgeom}
M.~R. Douglas,
\newblock {\em Topics in D-geometry},
\newblock Class. Quant. Grav. {\bf 17} (2000) 1057--1070, hep-th/9910170.

\bibitem{AD:Dstab}
P.~S. Aspinwall and M.~R. Douglas,
\newblock {\em D-Brane Stability and Monodromy},
\newblock JHEP {\bf 05} (2002) 031, hep-th/0110071.

\bibitem{me:point}
P.~S. Aspinwall,
\newblock {\em A Point's Point of View of Stringy Geometry},
\newblock hep-th/0203111.

\bibitem{AKH:m0}
P.~S. Aspinwall, R.~P. Horja, and R.~L. Karp,
\newblock {\em Massless D-Branes on \CY\ Threefolds and Monodromy},
\newblock hep-th/0209161.

\bibitem{DDG:wrap}
D.-E. Diaconescu, M.~R. Douglas, and J.~Gomis,
\newblock {\em Fractional Branes and Wrapped Branes},
\newblock JHEP {\bf 02} (1998) 013, hep-th/9712230.

\bibitem{Fiol:suN}
B.~Fiol,
\newblock {\em The BPS Spectrum of $N=2$ $\SU(N)$ SYM and Parton Branes},
\newblock hep-th/0012079.

\bibitem{me:en3g}
P.~S. Aspinwall,
\newblock {\em Enhanced Gauge Symmetries and Calabi--Yau Threefolds},
\newblock Phys. Lett. {\bf B371} (1996) 231--237, hep-th/9511171.

\bibitem{KMP:enhg}
S.~Katz, D.~R. Morrison, and M.~R. Plesser,
\newblock {\em Enhanced Gauge Symmetry in Type II String Theory},
\newblock Nucl. Phys. {\bf B477} (1996) 105--140, hep-th/9601108.

\bibitem{me:lK3}
P.~S. Aspinwall,
\newblock {\em K3 Surfaces and String Duality},
\newblock in C.~Efthimiou and B.~Greene, editors, ``Fields, Strings and
  Duality, TASI 1996'', pages 421--540, World Scientific, 1997,
\newblock hep-th/9611137.

\bibitem{CDFKM:I}
P.~Candelas et~al.,
\newblock {\em Mirror Symmetry for Two Parameter Models --- I},
\newblock Nucl. Phys. {\bf B416} (1994) 481--562, hep-th/9308083.

\bibitem{Horj:EZ}
P.~Horja,
\newblock {\em Derived Category Automorphisms from Mirror Symmetry},
\newblock math.AG/0103231.

\bibitem{HKTY:}
S.~Hosono, A.~Klemm, S.~Theisen, and S.-T. Yau,
\newblock {\em Mirror Symmetry, Mirror Map and Applications to Calabi--Yau
  Hypersurfaces},
\newblock Commun. Math. Phys. {\bf 167} (1995) 301--350, hep-th/9308122.

\bibitem{Thom:DCg}
R.~P. Thomas,
\newblock {\em Derived Categories for the Working Mathematician},
\newblock math.AG/0001045.

\bibitem{Doug:S01}
M.~R. Douglas,
\newblock {\em D-Branes and N=1 Supersymmetry},
\newblock hep-th/0105014.

\bibitem{Fzdn:msline}
A.~Fayyazuddin,
\newblock {\em Some Comments on $N=2$ Supersymmetric Yang--Mills},
\newblock Mod. Phys. Lett. {\bf A10} (1995) 2703--2708, hep-th/9504120.

\bibitem{AFS:msline}
P.~C. Argyres, A.~E. Faraggi, and A.~D. Shapere,
\newblock {\em Curves of Marginal Stability in $N=2$ Super-QCD},
\newblock in I.~Bars et~al., editors, ``Future Perspectives in String Theory,
  Strings '95'', Univ. Southern California, 1996,
\newblock hep-th/9505190.

\bibitem{BBS:5b}
K.~Becker, M.~Becker, and A.~Strominger,
\newblock {\em Five-branes, Membranes and Nonperturbative String Theory},
\newblock Nucl. Phys. {\bf B456} (1995) 130--152, hep-th/9507158.

\bibitem{OOY:Dm}
H.~Ooguri, Y.~Oz, and Z.~Yin,
\newblock {\em D-branes on Calabi-Yau Spaces and their Mirrors},
\newblock Nucl. Phys. {\bf B477} (1996) 407--430, hep-th/9606112.

\bibitem{Grot:P1}
A.~Grothendieck,
\newblock {\em Sur le classification de fibr{\'e}s holomorphes sur le
  sph{\`e}re de Riemann},
\newblock Am. J. Math. {\bf 79} (1957) 121--138.

\bibitem{MM:K}
R.~Minasian and G.~Moore,
\newblock {\em K-Theory and Ramond-Ramond Charge},
\newblock J. High Energy Phys. {\bf 11} (1997) 002, hep-th/9710230.

\bibitem{FW:D}
D.~S. Freed and E.~Witten,
\newblock {\em Anomalies in String Theory with D-branes},
\newblock hep-th/9907189.

\bibitem{HS:hom}
P.~Hilton and U.~Stammbach,
\newblock {\em A Course in Homological Algebra},
\newblock Springer-Verlag, 1970.

\bibitem{Friedm:algsurf}
R.~Friedman,
\newblock {\em Algebraic Surfaces and Holomorphic Vector Bundles},
\newblock Springer-Verlag, 1998.

\bibitem{Brig:flop}
T.~Bridgeland,
\newblock {\em Flops and Derived Categories},
\newblock math.AG/0009053.

\bibitem{ST:braid}
P.~Seidel and R.~P. Thomas,
\newblock {\em Braid Groups Actions on Derived Categories of Coherent Sheaves},
\newblock Duke Math. J. {\bf 108} (2001) 37--108, math.AG/0001043.

\end{thebibliography}

\end{document}